# *The search for longitude: Preliminary insights from a 17th Century Dutch perspective*

Richard de Grijs
*Kavli Institute for Astronomy & Astrophysics, Peking University, China*

**Abstract.** In the 17th Century, the Dutch Republic played an important role in the scientific revolution. Much of the correspondence among contemporary scientists and their associates is now digitally available through the *ePistolarium* webtool, allowing current scientists and historians unfettered access to transcriptions of some 20,000 letters from the Dutch Golden Age. This wealth of information offers unprecedented insights in the involvement of 17th Century thinkers in the scientific issues of the day, including descriptions of their efforts in developing methods to accurately determine longitude at sea. Unsurprisingly, the body of correspondence referring to this latter aspect is largely dominated by letters involving Christiaan Huygens. However, in addition to the scientific achievements reported on, we also get an unparalleled and fascinating view of the personalities involved.

## 1. Scientific dissemination during the Dutch 'Golden Age'

The 17th Century is regarded as the 'Golden Age' in the history of the Netherlands. The open, tolerant and transparent conditions in the 17th Century Dutch Republic – at the time known as the Republic of the Seven United Netherlands – allowed the nation to play a pivotal role in the international network of humanists and scholars before and during the 'scientific revolution'. The country had just declared its independence from the declining Spanish empire, in 1581, and its citizens had fully embraced the spirit of the European Enlightenment – more fully so than any other European nation at the time. The economist Dasgupta[1] has pointed out eloquently that this period was revolutionary precisely because "it created institutions that enabled the production, dissemination, and use of knowledge ... to be transferred from *small elites* to the *public at large*." Indeed, the global trade network established by this predominantly seafaring nation, the resulting prosperity and its political system's relative tolerance made the Dutch Republic a refuge for intellectuals, which in turn facilitated a dynamic intellectual atmosphere and free exchange of new knowledge in fields as diverse as philology, natural philosophy and natural history. As a consequence, intellectuals from all over Europe visited the Dutch Republic, which was then considered the 'storehouse of the intellectual world'.

However, in order for this dynamic intellectual atmosphere to thrive, scientific knowledge had to be disseminated somehow among the educated members of society. Given that the first scientific journal was not established until the 1660s[2], the key questions addressed by scholars focusing on this field include, *How did the 17th Century scientific information system work?* and *How were new*

---

[1] Dasgupta, P., 2007, *Economics: A Very Short Introduction*, Oxford, UK: Oxford University Press
[2] The first issue of the *Philosophical Transactions of the Royal Society of London* was published upon having been granted a Royal charter by King Charles II on 6 March 1665 (cf. http://rstl.royalsocietypublishing.org).

*scientific information and insights assimilated, distributed and eventually broadly established in the educated community?* Indeed, as we will see, the primary means of communication among the educated intellectuals consisted of a prolific exchange of letters – which, in turn, were often copied by their recipients (or their servants) and shared with other members of the educated society if and when relevant.

## 2. Source material

In this context, it is particularly interesting to note that the Dutch *Huygens ING* institute recently launched an online, virtual research environment, *ePistolarium*[3], under the umbrella of the 'Circulation of Knowledge and Learned Practices in the Seventeenth-Century Dutch Republic' (CKCC) project, which is freely accessible to anyone with an Internet connection. The *ePistolarium* data base contains the full text and metadata of some 20,000 'geleerdenbrieven' (scholarly letters) received and sent by nine 17th Century leading intellectuals who were based in the Dutch Republic during some or all of their life, including René Descartes and the father-and-son team Constantijn and Christiaan Huygens.

Given the fascinating combination of a nation that was very much outward looking, as shown by its extensive trade network around the world as well as its attraction to disenfranchised or persecuted scholars fleeing the censorship found elsewhere in Europe around the turn of the 17th Century, and the thriving intellectual environment in which some of the highest-profile contemporary scholars had established themselves, I became interested in perusing the *ePistolarium* correspondence to explore the role Dutch and Dutch Republic-based scholars had played in the development of instruments and technology that would allow suitably trained sailors to accurately and precisely determine their position, in particular the longitude, while at sea. This was particularly important for crews on the large fleet of Dutch merchant ships, because Spanish ports were off limits to Dutch Republic-flagged vessels. One of the key players among the commercial shipping giants was the *Vereenigde Oostindische Compagnie* (VOC; the Dutch East India Company), which sent its merchants to the far corners of the planet in pursuit of new trade links and rich returns from rare commodities brought back from those far-flung locales.

It was thus of the utmost importance to have access to reliable maps and charts, which were at the time considered state secrets. Exploration of new and strange lands was an integral aspect of these journeys, satisfying the new nation's unbridled intellectual curiosity. In the words of Carl Sagan[4], "the Middle Ages had ended; the Enlightenment had begun." Determination of one's latitude is fairly straightforward: one would simply need to measure the Sun's meridian altitude, corrected for the time of year, or note the southernmost stars visible at night. Longitude determination, on the other hand, required precise time keeping. If a sufficiently accurate shipboard time piece could be constructed, which would keep the time at one's home port, measuring the rising and setting times of known stars (which would enable determination of the local time) would directly lead to the determination of one's longitude by calculation of the difference between the local time and that at the home port. But the limited accuracy of existing, contemporary time pieces is precisely where the main problems arose. And this is, indeed, where Dutch scientists made a major impact at the time, particularly through the efforts of Christiaan Huygens.

---

[3] http://ckcc.huygens.knaw.nl/epistolarium/
[4] Sagan, C., 1980, *Cosmos* (Episode 6), PBS (USA)

I thus proceeded to search the *ePistolarium* data base extensively for terms that were broadly related to the development of accurate time pieces and efforts to use the latter to determine reliable positions at sea. Table 1 provides an overview of the search terms used and the numbers of letters returned. In addition to this primary source material, I also perused Christiaan Huygens' *Oeuvres Complètes*, based on both an electronic version of the originally published volumes[5] and a helpful translation into modern Dutch.[6] At the present time, I have only been able to reach preliminary conclusions about the involvement of and contributions by Dutch (and Dutch Republic-based) scholars and scientists to the development of technology allowing one to determine longitude at sea. This is what is presented in this contribution. Of the 355 letters I selected for further study, 241 (68%), 88 (25%), 18 (5%) and 8 (2%) were written in Renaissance French, 17th Century Dutch, Latin and contemporary English, respectively. (Except for the letters in Latin, for which convenient translations into modern Dutch are available, I am proficient at various levels in the other languages.)

*Table 1:* Search terms and numbers of letters

| *longitude* (45) | *latitude* (31) | *"map"* | *watch* (9) |
|---|---|---|---|
| - lengte (137) | - breedte (12) | - kaarte (2) | - horloge (101) |
| - lenkte (3) | - brete (1) | - caert (7) | - horloges (64) |
| - lengde (12) | - breete (54) | *horologie* (35) | - horologe (226) |
| *meridian* (3) | - breette (9) | - horologies (15) | - horologes (159) |
| - meridiaan (10) | | - horologien (48) | |

Figure 1 shows the distribution of the 355 letters used in this study throughout the 17th Century. The top panel shows (in red) the extent to which Christiaan Huygens was involved: 321 (90%) of the 355 letters originated from or were addressed to him, thus showing once more the importance of this pivotal scientist in the history of the Dutch Republic. The bottom panel includes only the 321 letters received or sent by Christiaan Huygens. He exchanged more than 20 letters with each of five of his contemporaries, for four of whom the distribution over time is shown in this bottom panel. (The fifth correspondent was his brother, Constantijn Jr.) One can discern two clear periods in which Huygens corresponded most intensely with his contemporaries, roughly centred on 1660 and 1690. These peaks include correspondence on a variety of topics, but the main subjects involve Huygens' invention and further development of the pendulum clock and the first few sea trials of the improved time pieces, respectively. I will return to these important milestones below.

---

[5] https://archive.org/details/oeuvrescomplte04huyg
[6] http://adcs.home.xs4all.nl/Huygens/

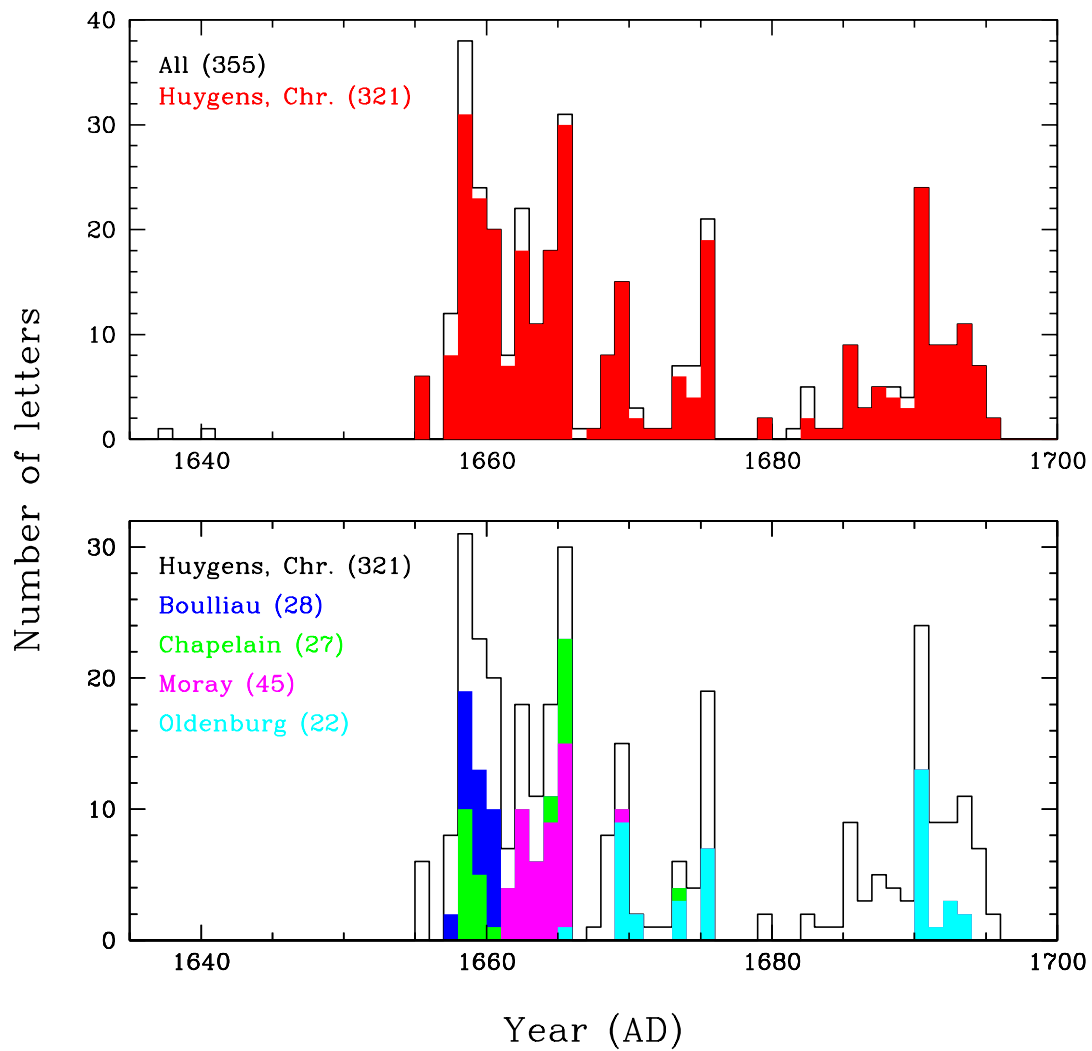

**Figure 1:** Body of letters from the *ePistolarium* data base related to the technological development of a method to accurately determine longitude at sea. Numbers in brackets refer to the numbers of letters identified. *Top*: Involvement of Christiaan Huygens in the body of correspondence. *Bottom*: Distribution of Huygens' letters and contributions by the four most prolific correspondents.

## 3. Inspired by Galileo Galilei?

The first time the need to determine accurate positions at sea is mentioned in the *ePistolarium* data base occurs in a letter[7] dated 13 April 1637 from Constantijn Huygens (hereafter Huygens Sr.) to Élie Diodati, a Swiss lawyer from Geneva, Switzerland, and a supported of Galileo Galilei. Diodati was based in Paris (France), where he helped Galilei to publish his manuscripts, given that the latter was unable to do so in his native Italy because of censorship imposed by the Catholic Church. Huygens Sr. refers in his letter to a proposal by Galilei to develop a device to determine longitude at sea, which would require stable operation of time pieces, in order to accurately determine the ephemerides, "as on solid ground". In his response, dated 28 February 1640[8], almost three years later, Diodati updates Huygens Sr. as regards a number of events that have happened in the mean time. He says that Galilei's proposal was submitted to the "Messeigneurs les Éstats Generaux", but that it has been affected by several interruptions, including the demise of all four commissaires who had been tasked with researching this proposal. As such, he asks Huygens Sr. in flowery and very flattering language to take on this task on behalf of Galilei, although I have not uncovered any evidence that Huygens Sr. made any progress in this respect.

---

[7] 1637-04-13: Huygens, Constantijn – Diodati, Élie; *De Briefwisseling van Constantijn Huygens* (http://resources.huygens.knaw.nl/retroboeken/huygens/#page=0&source=19; Rijks Geschiedkundige Publicatiën; Worp, J. A.), Martinus Nijhof: 's Gravenhage, letter No. 1542

Following this early exchange of letters, the next time this issue is discussed in the corpus of 17th Century letters we have access to is 15 years later, in 1655. The young, 26 year-old Christiaan Huygens (henceforth simply referred to as 'Huygens') is tasked by the Staten Generaal of the Dutch Republic, i.e., the national government, to assess a proposed invention by Johannes Placentinus, a professor of mathematics from Frankfurt an der Oder in Prussia (on the present-day border between Germany and Poland), for the purposes of patenting the idea. Placentinus' invention purports to allow the "determination of East and West" using observations of the Moon. For Huygens to be able to reach a decision on the suitability of Placentinus' method, in March 1655 he writes[9] to the scholar and preacher Andreas Colvius from Dordrecht (in the Dutch Republic), "… I expect, in turn, that you will send me manuscripts regarding the determination of longitude and whatever else you own from Galilei's legacy."

Huygens' assessment of Placentinus invention is that the proposed method is wholly unusable since it violates basic astronomical principles, including the assumption that the Moon would traverse 15 degrees on the sky in an hour (just like the Sun). This is indeed incorrect. As such, Huygens refers to the proposed invention as 'Placentinus' nonsense'[10]. Although most likely unintentional, this shows a character trait that comes back time and again in his letters: while he tends to be very polite to his addressees, he clearly is a man who does not suffer (those he considers) fools easily nor does he have much patience for those he believes to be lacking in education or insight. Perusing the vast number of letters left by Huygens, it becomes abundantly clear that he must have disliked one person in particular, notably the French philosopher René Descartes, whom he got to know fairly well as a family friend of his father's. The language Huygens uses in some of the letters in which he refers to Descartes would not pass peer review even at the present time…

In response to Huygens' negative assessment, the government's executive branch decides to consult Frans van Schooten, professor of mathematics at Leiden University (Dutch Republic), for a second opinion[11]. Huygens and van

---

[8] 1640-02-28: Diodati, Élie – Huygens, Constantijn; *Ibid.*, No. 2318
[9] 1655-03: Huygens, Christiaan – Colvius, Andreas; *Oeuvres Complètes de Christiaan Huygens,* Vol. 1, No. 217 (p. 322)
[10] 1655-03-04: Huygens, Christiaan – Staten Generaal; *Ibid.*, No. 214 (pp. 318–319)
[11] 1655-03-08: Staten Generaal – Huygens, Christiaan; *Ibid.*, No. 216 (pp. 320–321)

Schooten subsequently exchange numerous letters – all in Latin – between March 1655 and November 1656; van Schooten hence becomes *de facto* Huygens' mentor during the early phases of his career. Their exchange includes a copy of the *Opere di Galileo Galilei* (*Works of Galileo Galilei*), which is important in the context of our analysis of the Dutch efforts to determine longitude at sea, because the latter work included ideas about pendulums. Pendulum clocks will form the basis of the method developed by Huygens, who is in fact credited with the invention of the pendulum clock. Huygens himself concedes that he has seen Galilei's description, in a concept letter in Latin from 13 August 1657[12], although it should be noted that Galilei never actually completed the construction of a pendulum clock himself.

### 4. Opportunistic competition

On 12 January 1657, Huygens writes[13] to van Schooten that "one of these days, I invented a new type of construction for a time piece, which can be used to measure times so accurately that there is more than a little hope that this can be used to determine the longitude, at least as regards travel on the seas." Huygens' personal notes, collected in his *Adversaria*, imply that he invented the pendulum clock in the last days of December 1656. This is corroborated by a letter[14] to the French astronomer and mathematician Ismaël Boulliau from 26 December 1657, in which Huygens states that "yesterday, it was exactly a year ago that I made the first model."

His trick was to derive the pendulum's motion from that of the clock, while basing the clock's regularity on that of the pendulum. Huygens was, at least initially, keen to discuss his new invention with van Schooten and other contemporaries, including the French mathematician Claude Mylon, member of the famous *Académie Parisienne* (Parisian Academy), who told him in a letter dated 12 April 1657[15], "Your invention of a time piece is thought of very highly by all with whom I have discussed it. It will be even better if you can make sure that it is not affected by changes to either counterweight or spring. With such a watch you can derive a true equation of time, as required to determine the longitude." The equation of time Mylon refers to here is a correction that needs to be applied to the regular clock time to reflect the 'figure-of-eight' path the Sun traces on the sky throughout the year when we measure local noon based on determining the Sun's highest point on any given day. Almost simultaneously, in June 1657, Huygens contacted Samuel C. Kechelius van Hollenstein ('Kechel') to discuss the merits of pendulum clocks for accurate time keeping.[16,17]

---

[12] 1657-08-13: Huygens, Christiaan – unknown; *Ibid.*, Vol. 2, No. 400 (p. 46). This incomplete letter is found on the reverse of a concept of letter No. 399 (Huygens, Christiaan – de Sluse, René François)
[13] 1657-01-12: Huygens, Christiaan – van Schooten, Frans; *Ibid.*, No. 368 (p. 5)
[14] 1657-12-26: Huygens, Christiaan – Boulliau, Ismaël; *Ibid.,* No. 443 (p. 109)
[15] 1657-04-12: Mylon, Claude – Huygens, Christiaan; *Ibid.*, No. 382 (pp. 22–23)
[16] Although the addressee is not referred to by name, Kechel was at that time the only person known to have engaged in the astronomical eclipse (and other emphemeris) observations that were discussed. He had published a treatise on this topic: *Eygentlicke afbeeldinge der Drie Sonnen, dewelcke verschenen zijn, Ao. 1653 den 14/24 Jan. alhier binnen Leijden, ende op den Toren van de Academie waergenomen door Sam. Car. Kechel van Hollenstein* (1653).

However, he soon realized that he could patent his invention, thus leading to more cautious, almost secretive exchanges. Indeed, both the national and provincial governments, the Staten Generaal[18] and the Staten van Hollant ende Westvrieslant[19], awarded patents for the exclusive development, manufacturing and sale of pendulum clocks for a period of 21 years to Huygens and his master clock maker Salomon Coster of The Hague, in June and July 1657, respectively. The patents specified that these time pieces had certain characteristics, including a different movement mechanism from anything used previously, leading to more accurate time keeping that was not affected by changes in weather or mechanical faults.

To their significant distress, a year later – in August 1658 – the Staten General awarded[20] an almost identical patent to the Rotterdam city clock and watch maker, Simon Douw, "under false pretences". Huygens alleged[21] that Simon Douw secretly examined the inner workings of the Scheveningen public pendulum clock on 15 April 1658. He formally attested to this allegation and had this legalized by Hermanus de Coninck, notary public, in the presence of a well-respected witness. In this context, we get to see Huygens from a different side again. In letters to John Wallis[22], a high-profile English mathematician who is credited with introducing the symbol $\infty$ for infinity, to his cousin Willem Pieck[23] and his brother Lodewijk[24], as well as in a lively exchange with van Schooten[25], dated throughout September and October 1658, he calls Douw's behaviour "shameless and criminal" and the perpetrator himself "a stupid and shameless man". The controversy[26] and legal wranglings[27,28] continue well into 1659, without any clear resolution.

With Huygens' invention apparently out in the open, opportunistic competition causes him significant difficulties in monetizing his work. In June 1658, Ismaël Boulliau promises[29] Huygens to try to gain access to the Parisian market. However, Boulliau finds out[30] that Pierre Séguier – Chancellor of France from 1635 – "ne vouloit pas faire crier apres luy tous les maistres horologeurs de Paris": he did not want to alienate the master clock makers of Paris. However, the accuracy of Huygens' time piece required significant improvements before it could be used for the purpose it was originally invented for: longitude determination.

---

[17] 1657-06: Huygens, Christiaan – Kechel, Samuel; *Oeuvres Complètes de Christiaan Huygens,* Vol. 2, No. 392 (pp. 34–35)
[18] 1657-06-16: Staten Generaal – Coster, Salomon; *Ibid*., No. 525 (pp. 237–238)
[19] 1657-07-16: Staten van Hollant ende Westvrieslant – Coster, Salomon; *Ibid*., No. 526 (p. 239)
[20] 1658-08-08: Staten Generaal – Douw, Simon; *Ibid*., Nos 527, 528 (p. 240)
[21] *Horologium*, Appendix V (http://gallica.bnf.fr/ark:/12148/bpt6k778667/f86)
[22] 1658-09-06: Huygens, Christiaan – Wallis, John; *Oeuvres Complètes de Christiaan Huygens,* Vol. 2, No. 512 (pp. 210–211)
[23] 1658-10: Huygens, Christiaan – Pieck, Willem; *Ibid*., No. 532 (pp. 247–248)
[24] 1658-10-18: Huygens, Christiaan – Huygens, Lodewijk; *Ibid*., Vol. 22, p. 782
[25] 1658-10-04, -05, -13, -18: exchange between Huygens, Christiaan and van Schooten, Frans; *Ibid*., Vol. 2, Nos 523, 531, 534, 535 (pp. 235–236, 246, 249–250, 251, respectively)
[26] 1659-02-13: van Schooten, Frans – Huygens, Christiaan; *Ibid*., No. 587 (pp. 352–353)
[27] 1658-12-16: Raden van Hollandt, Zeelandt ende Vrieslandt – Staten van Hollant ende Westvrieslant; *Ibid*., No. 555 (p. 288)
[28] 1658-12-17: Staten van Hollant ende Westvrieslant – Douw, Simon; *Ibid*. No. 557 (pp. 291–292)
[29] 1658-06-13: Boulliau, Ismaël – Huygens, Christiaan; *Ibid*., No. 490 (pp. 183–184)
[30] 1658-06-21: Boulliau, Ismaël – Huygens, Christiaan; *Ibid*., No. 492 (pp. 185–186)

## 5. Refinements, obstacles and the long road to practical use

In addition to his numerous other endeavours, Huygens continued to improve his pendulum clock throughout the period 1658–1685. He embarked on a detailed assessment of the factors which may affect a clock's regularity at sea, although this was done without putting the clocks to the test. He considered what the effects on his clocks' regularity might be owing to changes in humidity and temperature, such as those one might encounter when sailing from temperate to tropical climates and back, as well as those due to changes in counterweight, pendulum length or spring balance, including their dependence on geographical location (i.e., under different gravitational strengths due to the slightly flattened shape of the Earth and because of the ship's motion and the effects of the Coriolis force, which takes into account the Earth's rotation). Meanwhile, towards the end of this period, the Dutch East India Company, VOC, took note of his efforts and issued a number of resolutions in support of taking one of Huygens' clocks on a journey to the Cape of Good Hope, the southernmost point of the African continent.

At long last, on 3 October 1685 Huygens writes[31] to Johannes Hudde, Mayor of Amsterdam, governor of the VOC and a mathematician by training, to announce the first test on the open water of the Zuyderzee, the former inland sea in the centre of the present-day Netherlands. The sea trial would take place on 11 October 1685, using two time pieces. However, because of a severe storm, the ship's captain is forced to abandon the journey, fearing damage to his sails. Nevertheless, and despite the rough conditions, one of the clocks continues to run smoothly.

The first real sea trial will thus need to be done during the first actual, VOC-sanctioned journey to the Cape. This first journey commences on 24 May 1686, arriving at the Cape on 26 September of that year. The return journey started on 20 April 1687, but the notes of Huygens' envoy, Thomas Helder, do not begin until 25 May[32]. Helder continues taking measurements of the clock's performance until 24 August 1687. He dies shortly afterwards. Figure 2 shows the journey taken by the '*Alcmaer*', on board of which Helder carried out his measurements, as well as the differences among the different methods used to determine the ship's position.

Since this first test was felt to have left too many open questions as regards the precision of the resulting longitudes[33], on 8 August 1689 the VOC issued a resolution to undertake a second sea trial. This time, Johannes de Graaff would sail on board a VOC ship (the *Alcmaer*) to the Cape. Huygens and de Graaff engage in extensive correspondence aimed at ensuring that the test would go well and that de Graaff is fully prepared for any unforeseen circumstances. The second sea trial runs from 28 December 1690 until 27 October 1692; on the outward journey, the ship arrived at the Cape on 3 June 1691. Upon having returned from the Cape and having filed the final report[34], Huygens and de Volder engage in a heated correspondence, driven by a fundamental difference in opinion between both men as regards the accuracy of the pendulum clocks provided by Huygens.

---

[31] 1685-10-03: Huygens, Christiaan – Hudde, Johannes; *Ibid.*, Vol. 9, No. 2401 (pp. 30–31)
[32] 1687-10-03: Huygens, Christiaan – de Graef, Abraham; *Ibid.*, No. 2488 (pp. 222–223)
[33] 1689-07-22: de Volder, Burchard – VOC; *Ibid.*, No. 2547 (pp. 339–343)

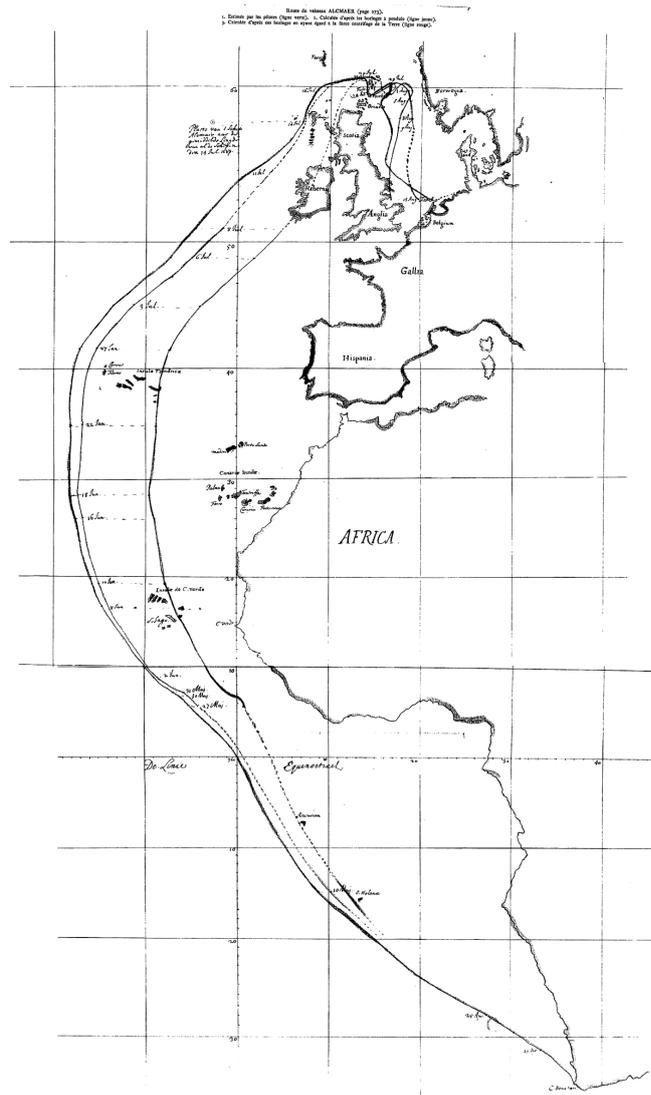

**Figure 2:** Route taken by the *Alcmaer*. The route determined by the ship's navigator is indicated by the westernmost curve, while that based on raw timing measurements is represented by the easternmost curve. Upon correction of the latter for the effects of the Earth's rotation, and recalibration of the ship's position along the way using known benchmarks, Helder and Huygens arrived at the curve closest to but slightly to the east of the westernmost route. *(Credit: Oeuvres Complètes de Christiaan Huygens, Vol. 9, No. 2519, pp. 272–291; 24 April 1688.)*

The accuracy required for the trial to be deemed successful had been clearly communicated by the VOC. Initially, a VOC resolution of 31 December 1682 called for an accuracy of better than 1 s deviation per 24 hours, although this was later (VOC resolution of 28 April 1684) relaxed to a required performance of better than 2 s per 24 hours. In a letter dated 24 March 1693, Huygens comments[35] to de Volder on the accuracy of the observations taken during the second journey; he concludes that a deviation of 10' 51 s was incurred over a period of 6 days. Although this was within the VOC's requisite level of accuracy, this would leave the Canary Island of Tenerife ~2.5° off from contemporary map measurements.

---

[34] 1692-11-19: de Graaff, Johannes – Huygens, Christiaan; *Ibid.*, Vol. 10, No. 2774 (p. 341)

At the end of his life, Huygens appears to mellow somewhat. In another letter[36] to de Volder, of 19 April 1693, he concedes that de Graaff's measurements needed correction and that he himself had made mistakes. As a possible solution, Huygens suggests that measurements taken on both legs of the journey *at the same place* should be compared with the accurately known timing of Jupiter's satellites to provide the final proof. (Contemporary maps cannot be used as benchmark, because they are uncertain themselves.) This brings the efforts to achieve an accurate determination of longitude at sea full circle: use of Jupiter's moons, whose ephemerides act as a very accurate natural clock, was suggested from the outset[37]. Practical considerations called for more straightforward timing measurements, however.

Huygens passes away on 8 July 1695, before being able to achieve his goal of manufacturing a sufficiently accurate time piece for use at sea.

## Acknowledgements

I would like to thank Prof. Xiaochun Sun (Institute for the History of the Natural Sciences, Chinese Academy of Sciences) for his encouragement to embark on this fascinating historical journey. I sincerely thank him and the conference organizers for the invitation to present this work; thanks to your efforts, I have discovered a new community which made me feel very welcome! Clearly, this project has only just started, but because of the community's encouragement, I am strengthened to continue my exploration of the corpus of 17th Century letters.

---

[35] 1693-03-24: Huygens, Christiaan – de Volder, Burchard; *Ibid.*, No. 2798 (pp. 433–434)
[36] 1694-04-19: Huygens, Christiaan – de Volder, Burchard; *Ibid.*, Nos 2802, 2803 (pp. 442–444)
[37] 1658-09-24: Huygens, Christiaan – Hodierna, Giovanni Batista; *Ibid.*, Vol. 2, No. 518 (pp. 223–225)